\documentclass[aps,10pt,prd,twocolumn,preprintnumbers,amsmath,amssymb,nofootinbib,superscriptaddress,a4paper,showpacs]{revtex4-1}
\pdfoutput=1
\usepackage{xcolor}
\definecolor{rindou1}{rgb}{0.4431,0.2862,0.7960}
\definecolor{rindou2}{rgb}{0.0078,0.1215,0.4392}
\definecolor{lapis}{rgb}{0.0.0470,0.2941,0.5568}
\definecolor{emerald}{rgb}{0.31, 0.78, 0.47}
\definecolor{deepsaffron}{rgb}{1.0, 0.6, 0.2}
\definecolor{pinegreen}{rgb}{0.0, 0.47, 0.44}
\definecolor{majorelleblue}{rgb}{0.38, 0.31, 0.86}
\definecolor{jade}{rgb}{0.0, 0.66, 0.42}
\definecolor{teal}{rgb}{0.0, 0.5, 0.5}
\definecolor{darkcyan}{rgb}{0.0, 0.55, 0.55}
\definecolor{jazzberryjam}{rgb}{0.65, 0.04, 0.37}
\definecolor{electricviolet}{rgb}{0.56, 0.0, 1.0}
\definecolor{regalia}{rgb}{0.32, 0.18, 0.5}
\definecolor{burgundy}{rgb}{0.5, 0.0, 0.13}
\definecolor{indigo(web)}{rgb}{0.29, 0.0, 0.51}
\definecolor{cerise}{rgb}{0.87, 0.19, 0.39}
\definecolor{darkbyzantium}{rgb}{0.36, 0.22, 0.33}
\definecolor{darkscarlet}{rgb}{0.34, 0.01, 0.1}
\definecolor{tyrianpurple}{rgb}{0.4, 0.01, 0.24}
\definecolor{amaranth}{rgb}{0.9, 0.17, 0.31}
\definecolor{iris}{rgb}{0.35, 0.31, 0.81}
\definecolor{darkred}{rgb}{0.55, 0.0, 0.0}
\definecolor{slateblue}{rgb}{0.42, 0.35, 0.8}
\usepackage[T1]{fontenc}
\usepackage{datetime2}
\usepackage[english]{babel}
\usepackage[utf8]{inputenc}
\usepackage{graphicx}
\usepackage{color}
\usepackage{amsfonts,amsthm}
\usepackage{bm,bbm}
\usepackage{mathrsfs}
\usepackage{amsmath}
\usepackage{float}
\usepackage{pgf,tikz}
\usepackage{mathrsfs}
\usetikzlibrary{arrows}

\usepackage[colorlinks=true, linkcolor=burgundy, urlcolor=lapis, citecolor=lapis, anchorcolor=green]{hyperref}  

\begin{document}
\title{Geometric Recursion from Polytope Triangulations and Twisted Homology}
\author{Nikhil Kalyanapuram }
\email{nkalyanapuram@psu.edu}
\affiliation{Department of Physics, The Pennsylvania State University, University Park PA 16802, USA}
\affiliation{Institute for Gravitation and the Cosmos, Department of Physics, The Pennsylvania State University,University Park, PA 16892, USA}

%%%%%%%%%%%%%%%%%%%%%%%%%%%%%%%%%%%%%%%%%%

\begin{abstract}
A geometric approach to understanding recursion relations for scattering amplitudes is developed. We achieve this by studying intersection numbers of triangulated accordiohedra presented as hyperplane arrangements. The cancellation of spurious divergences is subsequently realized as a topological no-boundary condition.
\end{abstract}

%%%%%%%%%%%%%%%%%%%%%%%%%%%%%%%%%%%%%%%%%%

\maketitle

%%%%%%%%%%%%%%%%%%%%%%%%%%%%%%%%%%%%%%%%%%
\section{Introduction}\label{sec:intro}
In a number of quantum field theories of practical interest, such as pure Yang-Mills, calculations using Feynman diagrams at high multiplicity are often prohibitive. The inefficiency of the Feynman expansion owing to the profusion of algebraic and combinatorial complexity indicates often that a simpler picture should be at work, especially considering the simplicity of the final answers involved. Some of the earliest evidence in support of this was supplied by the recursion relations of Berends and Giele \cite{Berends:1987me}, followed a couple of decades later by the on shell recursion derived by Britto, Cachazo, Feng and Witten \cite{Britto:2004ap,Britto:2005fq}.

Since the development of these techniques, a variety of practical problems could be solved with relative ease. Applications and extensions have been made to gravitational theories \cite{Cachazo:2005ca,ArkaniHamed:2008yf,Bedford:2005yy,BjerrumBohr:2005jr,Mason:2009sa}, generic theories \cite{Cheung:2008dn,Cheung:2015cba,Cheung:2015ota}, string models \cite{Boels:2008fc,Cheung:2010vn} and one loop integrands \cite{Bern:2005hs}. For the special case of Yang-Mills theory with maximal supersymmetry, the insight provided by the idea of recursion has led to the discovery of the integrand to all loop orders in the planar limit \cite{ArkaniHamed:2010kv,ArkaniHamed:2010gh,Bourjaily:2010wh}.

Attempts to find the geometric structures underlying amplitudes have also led to alternate interpretations of recursion relations. In particular, there is the realization that amplitudes are often volumes of certain \emph{polytopes} \cite{Hodges:2009hk,ArkaniHamed:2010gg,Arkani-Hamed:2017mur,Banerjee:2018tun,Raman:2019utu,Jagadale:2019byr,Aneesh:2019cvt,Srivastava:2020dly,Kojima:2020tox,John:2020jww}. This allows for recasting recursion relations as specific triangulations of the polytopes in question. In order to clarify this point, let us see how exactly this plays out.

In this approach to geometrizing amplitudes, known as the positive geometry program, the scattering amplitudes are recast in two dual pictures. On the one hand, certain polytopes (associahedra for planar $\phi^3$ for example) are realized as convex objects in kinematical space $\mathcal{K}_{n}$. A canonical form $\Omega_{\mathcal{P}}$, which can be defined uniquely for a given polytope $\mathcal{P}$ \cite{Arkani-Hamed:2017tmz}, is shown to encode the amplitude as a residue evaluated on the polytope,

\begin{equation}
    A = \mathrm{Res}_{\mathcal{P}}\Omega_{\mathcal{P}}.
\end{equation}
Alternatively, the same amplitude is realized as the volume of the dual polytope $\mathcal{P}^{*}$,

\begin{equation}\label{eq:2}
    \mathrm{Res}_{\mathcal{P}}\Omega_{\mathcal{P}} = \mathrm{Vol}(\mathcal{P}^{*}).
\end{equation}
Here, the recursion relation is realized as a triangulation of the dual polytope. It can be shown that a triangulation of the dual polytope leads to a recursive definition of the canonical form. 

The positive geometry program has traditionally focused on massless particles. More recently, the tools of twisted intersection theory \cite{yoshida1,yoshida2,yoshida3,Mimachi_2003,Mizera:2017cqs,Mizera:2017rqa,Mastrolia:2018uzb,Frellesvig:2019kgj,Frellesvig:2019uqt,Mizera:2019gea,Mizera:2020wdt} have been employed to provide a generalization of the positive geometry program to massive internal states and more generally for external particles with generic kinematics off the mass shell \cite{Kalyanapuram:2019nnf,Kalyanapuram:2020vil,Kalyanapuram:2020axt} in terms of polytopes known as accordiohedra. In light of these developments, one can ask if recursion relations can be understood by leveraging the ideas of positive geometries and intersection theory. It is our objective in this article to address this question.

The intersection theory of $n$ dimensional polytopes discussed in \cite{Kalyanapuram:2019nnf,Kalyanapuram:2020vil} entailed the realization of these polytopes as hyperplane arrangements $\mathbb{CP}^{n}-\lbrace{f_{1},...,f_{N}\rbrace}$, where the $f_{i}$'s are the $N$ facets of the polytopes, namely Stokes polytopes or accordiohedra \cite{Thibault:2017nnf}. The intersection theory is then defined by considering cycles (domains of intersection) and cocycles (top-dimensional differential forms), which are then \emph{twisted} by a connection defined by logarithmic singularities as these hyperplanes are approached.

Crucially, these works have focused on the intersection theory of cycles or cocycles with themselves. These can be computed by certain localization techniques (see \cite{Mizera:2017cqs,Mizera:2017rqa} for further details) and give rise to exotic quasi-string theory amplitudes and field theory amplitudes of scalar field theory with polynomial interactions respectively. What has not been discussed so far is the pairing between a cycle and cocycle associated to the accordiohedra $\mathcal{A}_{n}$ defined in $\mathbb{CP}^{n}$. As it turns out, these quantities serve to provide the ideal arena to discuss a geometric form of recursion relations.

In order to define this pairing, the data required is as follows. First, a function $F_{\alpha'}$ on $\mathbb{CP}^{n}$, with the property that it vanishes as the hyperplanes are approached. Secondly, one requires a differential form $\varphi_{\mathcal{A}_{n}}$, which diverges logarithmically near the hyperplanes. The pairing is then given by the integral,

\begin{equation}
    \langle{\mathcal{A}_{n}\otimes F_{\alpha'},\varphi_{\mathcal{A}_{n}}\rangle} = \int_{\mathcal{A}_{n}}F_{\alpha'}\varphi_{\mathcal{A}_{n}}.
\end{equation}
Here, $\alpha'$ is a regulator that is present for formal purposes (aspects of this were discussed in \cite{Kalyanapuram:2019nnf}). For suitable choices of this defining data, we will show that a large class of field theory amplitudes can be recast as the leading contributions of such intersection numbers in the limit of vanishing $\alpha'$.

In this light, we will then be able to realize recursion relations as a triangulation of the domain of integration $\mathcal{A}_{n}$, which is a cycle by virtue of the removal of hyperplanes. This gives a purely geometrical meaning to the idea of recursion in scattering amplitudes for a very large class of theories. Additionally, the meaning of spurious poles is also made transparent - they occur as a consequence of the individual simplices of the triangulations having nontrivial boundary. The cancellation of spurious singularities in the final results is then recast as a requirement that the sum over boundaries of the triangulating simplices add up to a trivial boundary.

\section{Cycles and Cocycles for Accordiohedra}\label{sec2:acc} 
We start by describing how cycles and cocycles are defined for an accordiohedron constructed out of a hyperplane arrangement. To keep the discussion short, we restrict ourselves to a relatively simple, yet nontrivial example, namely the case of $6$ particles scattering in $\phi^3 + \phi^4$ theory. Some of the tools tacitly used, namely the specific convex realizations of these polytopes have been discussed previously in \cite{Thibault:2017nnf,Kalyanapuram:2019nnf,Kalyanapuram:2020vil,Kalyanapuram:2020axt}.

Accordiohedra are defined for a collection $\mathcal{C}_{n}$ of dissections of an $n$-gon. For a given dissection $\mathcal{D}\in\mathcal{C}_{n}$, the accordiohedron is determined by comparing the elements of $\mathcal{C}_{n}$ to $\mathcal{D}$ and imposing a condition of \emph{compatibility}. $\mathcal{D}$-compatible dissections belonging to $\mathcal{C}_{n}$ then label the vertices of the accordiohedron, which can be shown to be a \emph{polytope}. Lower codimension faces are labelled by partial dissections.

\begin{figure}[H]
\centering
\includegraphics[width=0.3\textwidth]{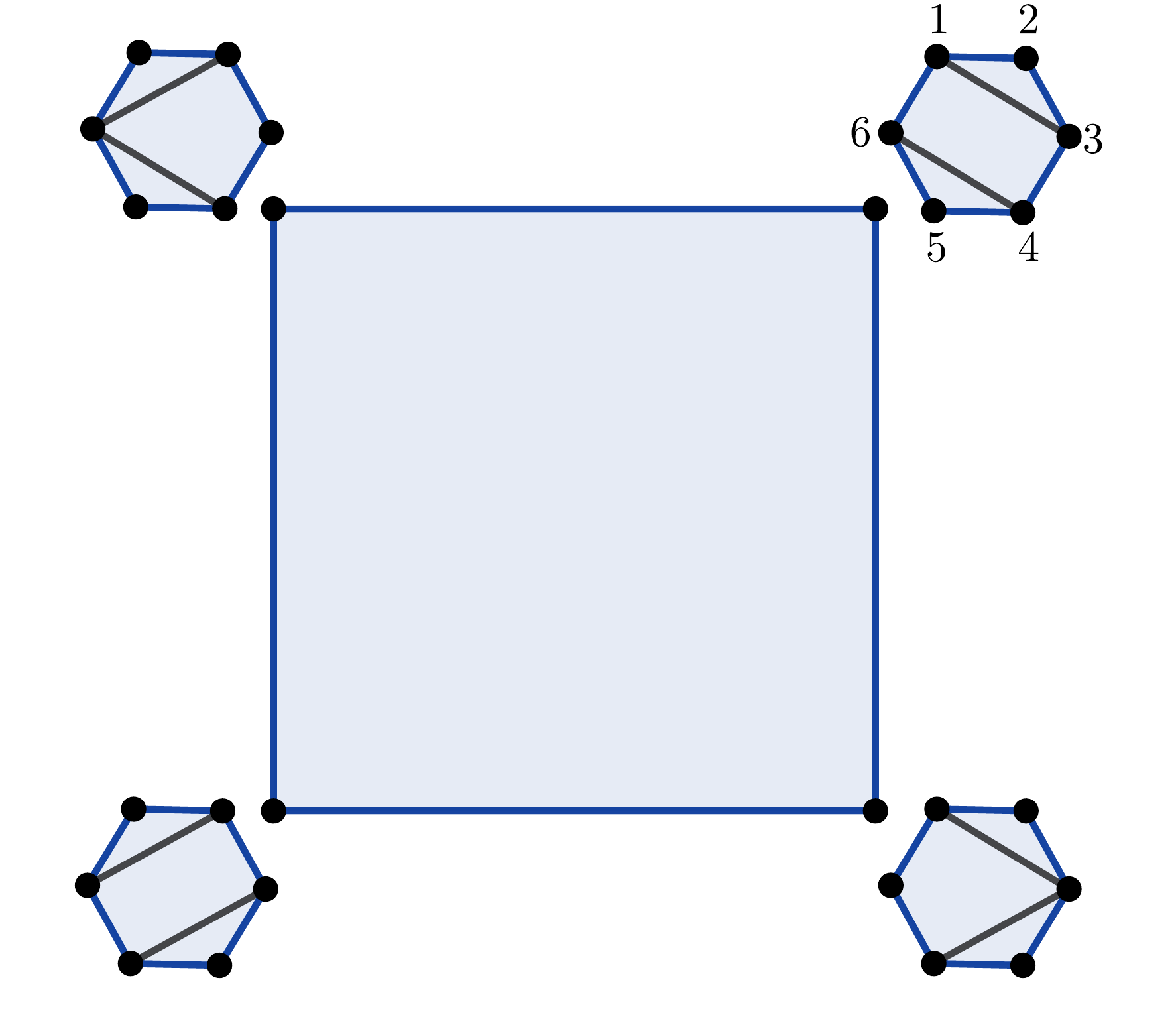}
\caption{Accordiohedron for the dissection $(13,46)$}\label{fig1}
\end{figure}

For our case, we need to focus on dissections of a hexagon with two triangulations and one quadrangulation. Here, we have four accordiohedra. For our purposes, we focus on the one that arises out of the dissection $(13,46)$. The accordiohedron is a square, with vertices being given as in figure \ref{fig1}. The reason it is a square is due to the fact that out of the $8$ possible dissections of the hexagon with two triangulations and one quadrangulation, only four are compatible with $(13,46)$.

Using now the rules for obtaining a convex realization, we can derive the following hyperplane arrangement for this polytope in $\mathbb{CP}^{2}$ as follows,

\begin{equation}
    \begin{aligned}
    &f_{1}: y+1 \;\; &f_{2}: x+1,\\
    &f_{3}: 1-y, \;\; &f_{4}: 1-x.
    \end{aligned}
\end{equation}
where we have used $(x,y)$ to denote inhomogeneous coordinates on $\mathbb{CP}^{2}$. It should be noted that the $f_{i}$ label the edges of the square clockwise, starting with the lower horizontal one (flanked by $(26,35)$ and $(13,35)$). In order to define a twisted homology and cohomology, we have to work on a space where the interior of this hyperplane arrangement can be realized as a \emph{cycle}, namely a domain of integration with vanishing boundary. This is possible if we define

\begin{equation}
    X_{(13,46)}:= \mathbb{CP}^{2}-\lbrace{f_1,f_2,f_3,f_4=0\rbrace}
\end{equation}
as our ambient space, always making sure to take into account the removal of the hyperplanes. Now the region bounded by the $f_{i}$s is trivially a cycle on $X_{(13,46)}$. 

Given this construction, a twist is a differential one-form that has logarithmic singularities as the hyperplanes are approached. The respective residues encode kinematical data. For our purposes, the desired twist is

\begin{equation}
\begin{aligned}
    &\omega_{(13,46)} = \\
    &(X_{35}-m^2)d\ln f_{1} + (X_{16}-m^2)d\ln f_{2}\\
    &(X_{46}-m^2)d\ln f_{3} + (X_{13}-m^2)d\ln f_{4}.
\end{aligned}
\end{equation}
Here, the variables $X_{ij}$ should not be confused with the space $X_{(13,46)}$; they simply represent planar Mandelstam variables defined by

\begin{equation}
    X_{ij} = (p_i+...+p_{j-1})^2.
\end{equation}
Additionally, we have used $m^2$ to denote the mass of the interacting scalars. Working at tree level, there is no issue, but these are bare masses if one wants to start handling loops.

Having these definitions now allows us to define the notion of twisted cycles and cocyles. A cycle is topologically a region of zero boundary, which in our case will be the region bounded by the hyperplanes. Now, a twisted cycle is a cycle that is \emph{loaded}, or in other words, multiplied by a function that vanishes as the hyperplanes are removed.\footnote{This definition is motivated by a study of hypergeometric functions. It turns out that it is this definition that allows for a recasting of hypergeometric integrals as intersection numbers.} For our purposes, we have the following twisted cycle,

\begin{equation}
    \mathrm{Acc}_{(13,46)} =([-1,1]\times[-1,1])\otimes \mathrm{KN}_{(13,46)}
\end{equation}
where,
\begin{equation}
    \mathrm{KN}_{(13,46)} = f_{1}^{i\alpha' (X_{35}-m^2)}\times ... \times f_{4}^{i\alpha'(X_{13}-m^2)}.
\end{equation}
Again here, as in the analysis done in \cite{Kalyanapuram:2019nnf}, an auxiliary parameter $\alpha'$ rears its head. It is introduced here mainly as a regulating parameter, and too much should not be thought of it. We will ultimately be interested in the final limit of $\alpha'\rightarrow 0$, so we won't discuss this particular point further, but will make some comments on this construction later in the article.

Now, twisted cocycles are differential forms that diverge logarithmically as the removed hyperplanes are approached. For our purposes, the most useful choice is,

\begin{equation}
\begin{aligned}
\varphi_{(13,46)} = &d\ln f_{1}\wedge d\ln f_{2}+d\ln f_2\wedge d\ln f_{3}\\+& d\ln f_3\wedge d\ln f_{4}d\ln f_{4}\wedge d\ln f_{1}.
\end{aligned}
\end{equation}
This is the standard representation for the canonical form of the square. The subscript simply indicates the reference dissection under consideration (a pedagogical review of a number of details involved in developing such forms and a twisted cohomology for accordiohedra can be found by the interested reader in \cite{Kalyanapuram:2020vil}.).

There are two things we now have to clarify before moving on - the sense in which these things are cycles and cocyles in the sense of a homology theory and how they give rise to intersection numbers. To do this, we observe that the operator

\begin{equation}
    \nabla_{(13,46)} = d + \omega \wedge
\end{equation}
is nilpotent. Accordingly, it defines a chain complex and a corresponding cohomology $H^{n}(X_{(13,46)},\nabla_{(13,46)})$ on the space $X_{(13,46)}$. It can be shown that the form $\varphi_{(13,46)}$ belongs to the highest cohomology group ($n=2$) of the chain complex. Additionally, it can be shown that the twisted cycle $\mathrm{Acc}_{(13,46)}$ is an element of the twisted homology. Further details and a number of worked out examples for associahedra can be found in \cite{Mizera:2017cqs}. 

With this, we have the twisted pairing of the cycle and cocyle as the integral,

\begin{equation}\label{eq12}
\begin{aligned}
    &\langle{\mathrm{Acc}_{(13,46)},\varphi_{(13,46)}\rangle} = \\
    &\int_{[-1,1]\times[-1,1]}(f_{1}^{\alpha' (X_{35}-m^2)}\times ... \times f_{4}^{\alpha'(X_{13}-m^2)})\varphi_{(13,46)}.
\end{aligned}
\end{equation}
Naturally, this integral depends on the parameter $\alpha'$. However, in the limit that this $\alpha'$ tends to zero, we claim that the leading contribution is the field theory amplitude coming from the twisted pairing of two accordiohedra,

\begin{equation}\label{eq13}
\begin{aligned}
   & \lim_{\alpha'\rightarrow 0} (\alpha')^2\langle{\mathrm{Acc}_{(13,46)},\varphi_{(13,46)}\rangle}\\
    =& \frac{1}{(X_{35}-m^2)(X_{16}-m^2)}+\frac{1}{(X_{16}-m^2)(X_{46}-m^2)}\\&+\frac{1}{(X_{13}-m^2)(X_{46}-m^2)}+\frac{1}{(X_{13}-m^2)(X_{35}-m^2)}.
\end{aligned}
\end{equation}
Proving this equation amounts to a rederivation of the localization formula proved in Appendix A of \cite{Mizera:2017cqs}. For completeness nowever, we note that this is proven by noting that the foregoing integral in equation (\emph{\ref{eq12}}) is evaluated for generic exponents by regularization on a Pochhammer contour. As the Pochhammer contour encircles a vertex comprehended by hyperplanes $f_{i}$ and $f_{j}$, it picks up a factor $(e^{2\pi i \alpha ' X_{i}}-1)^{-1}(e^{2\pi i \alpha ' X_{j}}-1)^{-1}$. In the limit of vanishing $\alpha'$, the denominator structure in equation (\emph{\ref{eq13}}) is recovered as these constitute the most divergent contributions. The $(2\pi i)$ factors are compensated for by contour integrals around the vertices. 

\section{Geometric Recursion via Triangulations}\label{sec3:geom}
Drawing on the ideas set forth to understand recursion in \cite{ArkaniHamed:2010gg,Arkani-Hamed:2017mur,Kojima:2020tox}, our focus now will be to derive recursion relations for generic amplitudes with polynomial interactions. To do this, we note two things. First, using intersection theory, amplitudes for a very large class of theories, including polynomial scalar field theories with masses and even some loop integrands can be derived \cite{Kalyanapuram:2020vil}.

Furthermore, as we illustrated with a simple example in the foregoing section, the cocycle-cocycle intersection numbers which gave rise to these amplitudes can actually be rederived as cycle-cocycle intersection numbers. The advantage here is that a cycle is a polytope, which can be triangulated, and this triangulation gives rise to the desired recursion. 

Let us be quantitative about this. For generality, suppose we have a hyperplane arrangement $f_1,...,f_{N}$ in $\mathbb{CP}^n$ bounding an accordiohedron $\mathcal{A}_{n}$, which is a cycle in the space $\mathbb{CP}^n$ with the hyperplanes removed. More precisely,

\begin{equation}\label{eq14}
    \partial \mathcal{A}_{n} = 0
\end{equation}
in $\mathbb{CP}^{n}-\lbrace{f_1,...,f_N\rbrace}$.

Now we consider a specific triangulation of this domain of integration into $m$ disjoint simplices, namely a decomposition,

\begin{equation}
    \mathcal{A}_{n} = \vec{\Delta}^{(1)}_{n}\cup ... \cup \vec{\Delta}^{(m)}_{n},
\end{equation}
where the arrow indicates that the simplices are \emph{oriented}. This is due to the fact that the no-boundary condition (\emph{\ref{eq14}}) place conditions on the boundaries on the simplices. In order to satisfy equation (\emph{\ref{eq14}}), we require,

\begin{equation}
    \sum_{i=1}^{m}\partial \vec{\Delta}^{(i)}_{n}=0.
\end{equation}

Having said this, the recursion relation for the amplitude simply becomes the following formula,

\begin{equation}\label{eq17}
    \begin{aligned}
    \langle{\mathcal{A}_{n}\otimes \mathrm{KN}_{\mathcal{D}},\varphi_{\mathcal{D}}\rangle} = \sum_{i=1}^{m}\langle{\vec{\Delta}^{(i)}_{n}\otimes \mathrm{KN}_{\mathcal{D}},\varphi_{\mathcal{D}}\rangle}.
    \end{aligned}
\end{equation}
We note here that we have used $\mathcal{D}$ to denote the particular dissection giving rise to the accordiohedron in question. The twisted pairing on the left side of the equation is defined in analogy to (\emph{\ref{eq13}}). 

A couple of things can be said about the foregoing formula. Firstly, the individual terms on the right side are \emph{not} genuine twisted pairings. Only their sum is an intersection number. Consequently, an evaluation of the individual terms should be expected to give spurious poles which cancel in the sum. As we will see in a specific example, these arise on account of the fact that the simplices have nontrivial boundary. Accordingly, the requirement that spurious poles cancel is just the realization at the level of the amplitude of the no-boundary condition (\emph{\ref{eq14}}).

We also note that equation (\emph{\ref{eq17}}) holds to all orders in the parameter $\alpha'$. Indeed, what we have just described would hold equally well as a recursion relation for string amplitudes described by intersection numbers, if $\alpha'$ is identified with the string tension. 

\subsection{How is this a recursion?}
What might not be clear at the outset by looking at formula (\ref{eq17}) is how this is a "recursion" relation. Indeed, it is by no means obvious that we are really recursing something to obtain the formula.

Clarifying this requires us to look back to the duality established by equation (\emph{\ref{eq:2}}). Here, we see a twofold description of scattering amplitudes in terms of positive geometries. Here, we have focused on the dual approach, dealing directly with the cycles (volumes) of the accordiohedron polytopes. We could instead have focused on the cocycle description (canonical forms). Here, we would have obtained a genuine recursion, via some kind of complex shift. While this duality has been well established for in the positive geometry story \cite{Arkani-Hamed:2017mur,John:2020jww}, a formal proof in the intersection theory story has yet to be supplied. This will serve as an important direction for future research.

In the next section, we will focus on a specific example to illustrate our result more effectively.

\section{Recursion in $\phi^3+\phi^4$ Theory}
Let us apply the result of section \ref{sec3:geom} to the case of the square accordiohedron corresponding to the dissection $(13,46)$ discussed in section \ref{sec2:acc}. The domain of integration is $[-1,1]\times[-1,1]$, which we split into two simplices as follows,

\begin{equation}
    \int_{-1}^{1} dx \int^{1}_{-1} dy = \int ^{1}_{-1}dy\int^{y}_{-1}dx + \int^{1}_{-1}dy\int^{1}_{y}dx. 
\end{equation}
We have written the right side of the equation in a manner that makes the specific orientations of the simplices manifest. To compute the intersection number this way, we note the following,

\begin{equation}\label{eq:19}
    \begin{aligned}
    \int_{-1}^{y}(1+x)^{a}(1-x)^{b} dx = 2^{a+b+1}\left(F_{1}(b,a)-F_{2}(y,b,a)\right)
    \end{aligned}
\end{equation}
and
\begin{equation}\label{eq:20}
    \begin{aligned}
   \int_{1}^{y}(1+x)^{a}(1-x)^{b} dx =2^{a+b+1}F_{2}(y;b,a)
    \end{aligned}
\end{equation}
where we have defined
\begin{equation}
    F_{1}(b,a) = B(b+1,a+1)
\end{equation}
and
\begin{equation}
    F_{2}(y,b,a) = B\left(\frac{1-y}{2};b+1,a+1\right)
\end{equation}
where $B(x_1;a_1,b_1)$ is the incomplete Beta function. 
In our discussion of the square accordiohedron, we encounter integrals of the following type,

\begin{equation}
    \begin{aligned}
    \int_{-1}^{1}dy\int_{-1}^{1}dx 
    (1+x)^{a}(1-x)^{b}(1+y)^{c}(1-y)^{d}.
    \end{aligned}
\end{equation}
Making use of equations (\emph{\ref{eq:19}}) and (\emph{\ref{eq:20}}), we can evaluate the first integral in this formula to obtain the following,

\begin{equation}
    \begin{aligned}
    &\int_{-1}^{1}dy (1+y)^{c}(1-y)^{d}\times\\
    &2^{1+a+b}\left(F_{1}(b,a)-F_{2}(y,b,a)+F_{2}(y,b,a)\right)
    \end{aligned}
\end{equation}
which reduces to

\begin{equation}
   2^{1+a+b}F_{1}(b,a) \int_{-1}^{1}dy (1+y)^{c}(1-y)^{d}.
\end{equation}
Here we note that the cancellation of the incomplete Beta function terms is the cancellation of spurious divergences. One can check (by somewhat tedious expansion) that these terms individually supply terms such as $(a+c)^{-1}$. Once we use this formula to compute the amplitude, we will note that this is indeed a spurious divergence, which quite gratifyingly drops out.

The $y$ integral can be carried out by simply taking the $y=1$ limit of equation (\emph{\ref{eq:19}}), to give us the final answer as,

\begin{equation}
    \begin{aligned}
    &=2^{2+a+b+c+d}\frac{\Gamma[1+a]\Gamma[1+b]\Gamma[1+c]\Gamma[1+d]}{\Gamma[2+a+b]\Gamma[2+c+d]}.
    \end{aligned}
\end{equation}
which we call $I_{(13,46)}[a,b,c,d]$.  

With this formula, we can proceed to evaluate the intersection number (\emph{\ref{eq12}}). The final answer devolves upon the evaluation of four terms,

\begin{equation}
    \begin{aligned}
    &I_{(13,46)}[Y^{\alpha'}_{35}-1,Y^{\alpha'}_{46},Y^{\alpha'}_{16}-1,Y^{\alpha'}_{13}]\\
    &I_{(13,46)}[Y^{\alpha'}_{35},Y^{\alpha'}_{46}-1,Y^{\alpha'}_{16}-1,Y^{\alpha'}_{13}]\\
    &I_{(13,46)}[Y^{\alpha'}_{35},Y^{\alpha'}_{46}-1,Y^{\alpha'}_{16},Y^{\alpha'}_{13}-1]\\
    &I_{(13,46)}[Y^{\alpha'}_{35}-1,Y^{\alpha'}_{46},Y^{\alpha'}_{16},Y^{\alpha'}_{13}-1].
    \end{aligned}
\end{equation}
where we have defined

\begin{equation}
    Y^{\alpha'}_{ij} = \alpha' (X_{ij}-m^2).
\end{equation}

The expansion of this formula can be carried out by use of the well known expansion of the Gamma function near $x=0$,
\begin{equation}
    \Gamma[\epsilon] = \frac{1}{\epsilon} + O(\epsilon^{0}).
\end{equation}

Using this, we have the following as the leading contribution to the amplitude as we take the limit of vanishing $\alpha'$

\begin{equation}
    \begin{aligned}
    &\frac{1}{\alpha'^2 (X_{35}-m^2)(X_{16}-m^2)}+\frac{1}{ \alpha'^2 (X_{46}-m^2)(X_{16}-m^2)}\\
   +& \frac{1}{ \alpha'^2 (X_{46}-m^2)(X_{13}-m^2)}+\frac{1}{ \alpha'^2 (X_{35}-m^2)(X_{13}-m^2)}.
    \end{aligned}
\end{equation}
which evidently reproduces the correct divergence structure of equation (\emph{\ref{eq13}}). Since we are working with real momenta, we had no trouble directly performing the integrals, which would otherwise have required a regularization using higher dimension Pochhammer contours.

Now we comment on the appearance of spurious poles. Observing that the evaluation on each triangulation gives denominators $(a+c)^{-1}$, for say $a =  \alpha' (X_{35}-m^2)$ and $c =  \alpha' (X_{13}-m^2)$, we see that this is indeed a spurious pole. Fortunately, we did not have to deal with them, as they nicely cancelled out when the contributions from the two triangulations were summed together.

Finally, we make some comments on how this formalism actually tells us how to perform this kind of recursion for theories with more generic numerator structure. For a familiar example, suppose we work with four-fermi theory. This is a theory with only quartic vertices, so it would be well described by accordiohedra. A square accordiohedron in this theory would describe $8$ particle scattering. Here, we can define the scattering form $\varphi$ by requiring that while it diverges logarithmically as a hyperplane is reached, the \emph{residue} near a vertex would not be $1$, but rather the numerical kinematical factor derived from the Feynman expansion.

\section{Discussion}
Our focus in this article has been to present a description of recursion relations in quantum field theory in a fashion that exploits modern geometric approaches to scattering amplitudes, namely the theory of intersection numbers and polytopes. Since we are able to leverage the fact that we can now describe quantum field theories with arbitrarily complicated interactions systematically using polytopes, the approach to recursion given here can be applied to nearly any theory of interest.

In \cite{Baadsgaard:2015voa,Baadsgaard:2015ifa,Baadsgaard:2015hia,Baadsgaard:2016fel}, among other things, rules were formulated for evaluating integrals that showed up in the moduli space formalism of Cachazo, He and Yuan \cite{Cachazo:2013gna,Cachazo:2013hca,Cachazo:2013iaa,Cachazo:2014nsa,Cachazo:2014xea,Cachazo:2015aol}. These were then used to provide a string-based model to describe scattering amplitudes for scalar theories with arbitrary interactions. It is hard not to suspect that this bears some relationship to the perspective outlined in \cite{Kalyanapuram:2020vil} and this work, the pursuing of which could be of value.

It should be emphasized that in this article, we have taken the right side of the duality expressed in (\emph{\ref{eq:2}}) as our approach to understanding recursion, by realizing recursion relations as triangulations of cycles. More traditional however is the point of view view indicated by the left side, in which the canonical form \emph{viz.} the amplitude itself is recursed by some kind of complex shift. The understanding of recursion relations applied to accordiohedra developed in the papers \cite{Kojima:2020tox,John:2020jww} follows this latter perspective. Investigating possible connections between these ideas and the techniques developed in this article would merit further research.

The fact that accordiohedra can be used to describe theories with \emph{arbitrarily} complicated interactions means that the formalism laid out here and in \cite{Kalyanapuram:2020vil} are likely to be especially well suited to handling theories that are of practical significance. It would be very interesting to direct future research towards applying these geometric notions to theories of direct experimental relevance, such as electroweak theory and theories with spontaneous symmetry breaking.

Lastly, the study of intersection theory has also shed new light recently on the phenomenon of the duality between color and kinematics \cite{Mizera:2019blq}. Exploring the implications of these ideas in the context of accordiohedra and the possibility of realizing recursion relations for amplitudes which can be expressed as double copies may also be interesting as a topic of further research.
%%%%%%%%%%%%%%%%%%%%%%%%%%%%%%%%%%%%%%%%%%

\section*{ACKNOWLEDGEMENTS} 
The author would like to thank Jacob Bourjaily for his sustained guidance and encouragement. The author also thanks Nima Arkani-Hamed and Sebastian Mizera for comments on the draft and Seyed Faroogh Moosavian for discussions. This project has been supported by an ERC Starting Grant (No. 757978) and a grant from the Villum Fonden (No. 15369).
%%%%%%%%%%%%%%%%%%%%%%%%%%%%%%%%%%%%%%%%%%

\section*{APPENDIX}
\begin{figure}[H]
\centering
\includegraphics[width=0.3\textwidth]{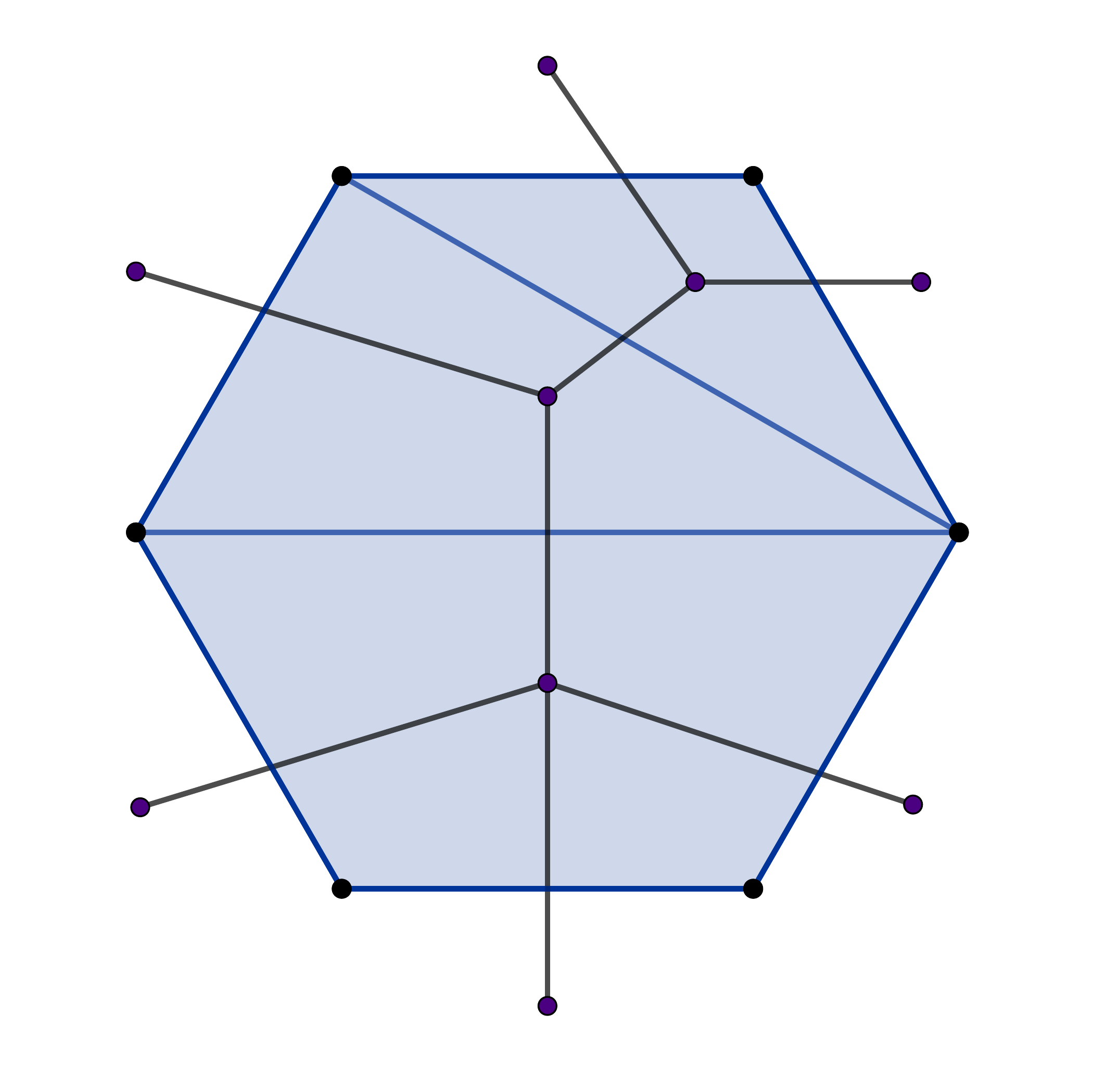}
\caption{Feynman Diagrams for $6$-particles in $\phi^4$.}\label{figa0}
\end{figure}

Let us briefly review some of the essential facts involved in the definition of accordiohedra. Accordiohedra are polytopes that generalize the associahedron - now understood to be the positive geometry of $\phi^3$ interactions - to generic scalar theories. In order to define these polytopes, we need to consider a collection $\mathcal{C}_{n}$ of dissections of an $n$-gon. Dissections of containing $p_3$ triangulations, $p_4$ quandrangulations etc. are in one-to-one duality between Feynman diagrams containing $p_3$ cubic vertices, $p_4$ quartic vertices and so on. A simple illustration of this fact is furnished in figure \ref{figa0}, representing a $6$ particle process.

\begin{figure}[H]
\centering
\includegraphics[width=0.4\textwidth]{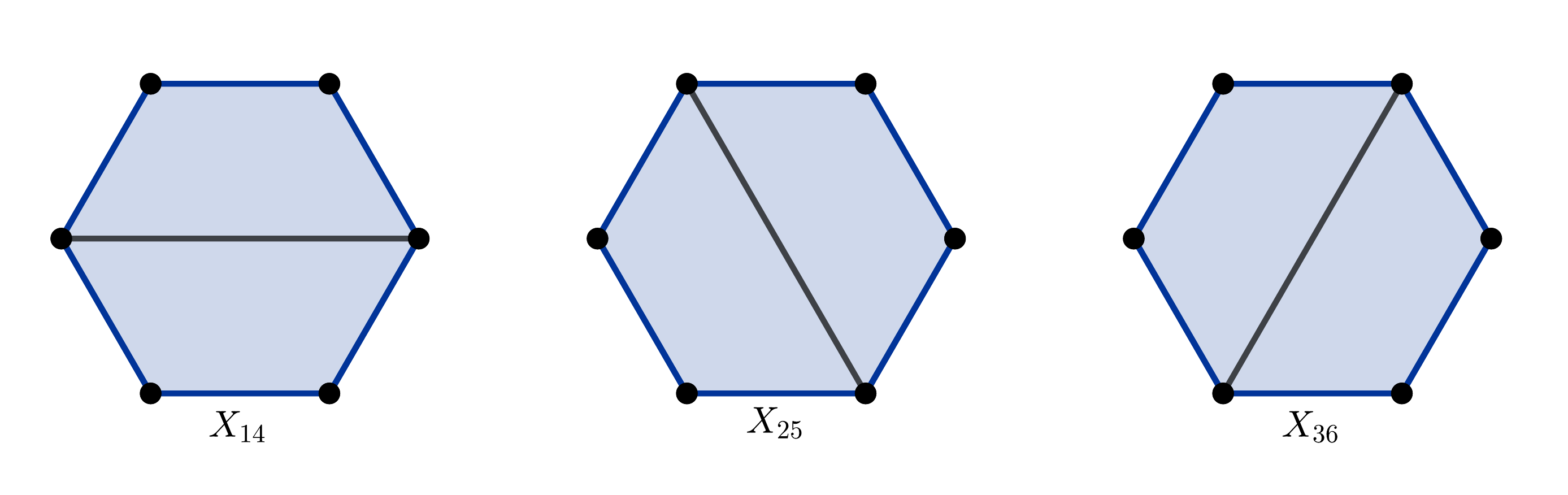}
\caption{Feynman Diagrams for $6$-particles in $\phi^4$.}\label{figa1}
\end{figure}

For purposes of concreteness, let us consider the simplest case, namely that of six particle scattering in $\phi^4$ theory. The planar Feynmn diagrams contributing at tree level are those with momentum transfer $X_{14} = s_{123}$, $X_{25}$ and $X_{36}$ as shown in figure \ref{figa1} 

Accordiohedra unlike associahedra are not unique. An accordiohedron is chosen by picking a particular dissection, let us choose $(14)$, and finding those dissections \emph{compatible} with it. These label the vertices of the accordiohedron, which can then be shown to be a convex polytope. This was the construction first applied to scattering amplitudes in a series of works by the author and collaborators \cite{Jagadale:2019byr,Kalyanapuram:2019nnf,Kalyanapuram:2020axt,Kalyanapuram:2020vil}. 

\begin{figure}[H]
\centering
\includegraphics[width=0.2\textwidth]{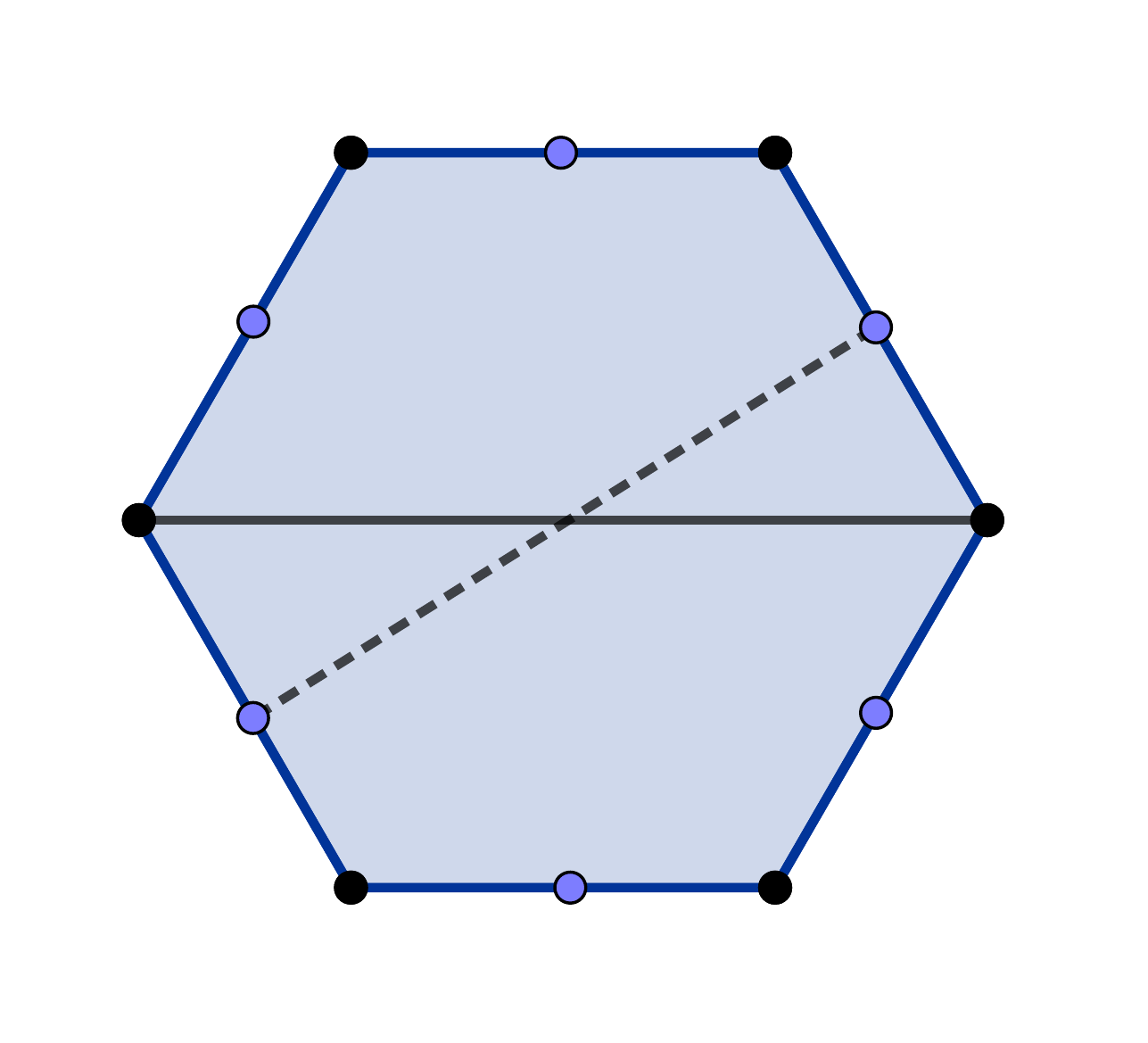}
\caption{$(36)$ drawn on the dual graph of $(14)$}\label{figa2}
\end{figure}

Let us now see how this compatibility condition works in practice. In order to see if a dissection $\mathcal{D}'$ is compatible with $\mathcal{D}$, each of its chords must be $\mathcal{D}$-accordion dissections. This means that when the chords of $\mathcal{D}'$ are drawn on the \emph{dual} of $\mathcal{D}$ (labelled by purple vertices in figure \ref{figa2}), the subgraph cut out must be connected. This is seen to be the case with $(36)$ drawn on $(14)$.

We can see from figure \ref{figa2} that the chord $(36)$ cuts the edges $(34)$ and $(61)$ as well as the chord $(14)$. The graph so formed is obviously connected. Accordingly, $(36)$ is compatible with $(14)$. The reader can verify as a quick exercise that $(14)$ is self-compatible while $(25)$ is not compatible with itself. 

This construction can be made arbitrarily complicated. In particular, we can do this for theories with polynomial interactions; there the dissections will be in terms of splitting an $n$ gon into triangle, quadrilaterals \emph{etc.} depending on whether we have interactions that are cubic, quartic and so on respectively. 
%%%%%%%%%%%%%
\bibliographystyle{utphys}
\bibliography{v1.bib}
%%%%%%%%%%%%%

\end{document}